\documentclass[twocolumn]{aastex631}

\newcommand{\rxc}{RXCJ2248-ID}

\newcommand{\iii}{~\textsc{iii}}

\newcommand{\mstar}{M$_*$}

\newcommand{\te}{$T_{\mathrm{e}}$}

\newcommand*{\msun}{\ensuremath{M_\odot\ }}

\newcommand*{\fesc}{\ensuremath{f_{esc}\ }}

\newcommand*{\Ha}{H\ensuremath{\alpha}\ }
\newcommand*{\Hb}{H\ensuremath{\beta}\ }

\newcommand*{\xOIII}{[O\,{\scshape iii}]\,\ensuremath{\lambda5007 \AA}}

\newcommand*{\sigmaSFR}{\ensuremath{\Sigma_{\rm SFR}\ }}
\newcommand*{\sigmaMass}{\ensuremath{\Sigma_{\rm M \ast}\ }}

\newcommand*{\OIII}{[O\,{\scshape iii}]}
\newcommand*{\NeIII}{[Ne\,{\scshape iii}]}
\newcommand*{\OII}{[O\,{\scshape ii}]}

\def\flyf{\ifmmode f_{\rm Lyf} \else $f_{\rm Lyf}$\fi}
\def\pz{\ifmmode P(z) \else $P(z)$\fi}
\def\ki2{\ifmmode \chi^2 \else $\chi^2$\fi}
\def\zphot{\ifmmode z_{\rm phot} \else $z_{\rm phot}$\fi}

\newcommand{\xphot}{\ifmmode x_\gamma \else $v_\gamma$\fi}
\newcommand{\xobs}{\ifmmode x_{\rm obs} \else $x_{\rm obs}$\fi}
\newcommand{\xcmf}{\ifmmode x_{\rm CMF} \else $x_{\rm CMF}$\fi}
\newcommand{\vexp}{\ifmmode V_{\rm exp} \else $V_{\rm exp}$\fi}
\newcommand{\nh}{\ifmmode N_{\rm HI} \else $N_{\rm HI}$\fi}
\newcommand{\dv}{\ifmmode \Delta v({\rm em-abs}) \else $\Delta v({\rm em}-{\rm abs})$\fi}

\def\frellya{\ifmmode f^{\rm rel}_{\rm{Ly}\alpha} \else $f^{\rm rel}_{\rm{Ly}\alpha}$\fi}

\newcommand{\muv}{\ifmmode M_{1500} \else $M_{1500}$\fi}
\newcommand{\auv}{\ifmmode A_{\rm UV} \else $A_{\rm UV}$\fi}
\newcommand{\luv}{\ifmmode L_{\rm UV} \else $L_{\rm UV}$\fi}
\newcommand{\lir}{\ifmmode L_{\rm IR} \else $L_{\rm IR}$\fi}
\newcommand{\lbol}{\ifmmode L_{\rm bol} \else $L_{\rm bol}$\fi}
\newcommand{\liruv}{\ifmmode L_{\rm IR+UV} \else $L_{\rm IR+UV}$\fi}
\newcommand{\liroveruv}{\ifmmode L_{\rm IR}/L_{\rm UV} \else $L_{\rm IR}/L_{\rm UV}$\fi}
\newcommand{\nlyc}{\ifmmode N_{\rm Lyc} \else $N_{\rm Lyc} $\fi}
\newcommand{\rholyc}{\ifmmode \rho_{\rm Lyc} \else $\rho_{\rm Lyc} $\fi}
\newcommand{\chion}{\ifmmode \xi_{\rm ion} \else $\xi_{\rm ion}$\fi}
\newcommand{\chioncorr}{\ifmmode \xi_{\rm ion}^0 \else $\xi_{\rm ion}^0$\fi}

\shorttitle{Physical properties of GHZ2 from combined NIRSPec and MIRI observations}
\shortauthors{Calabr\`o et al.}
\usepackage[autopunct=true]{csquotes}

\usepackage{amsmath}
\usepackage{graphicx}
\usepackage{natbib}
\usepackage{scalerel}
\usepackage{makecell}
\usepackage{multirow}
\usepackage{txfonts}

\usepackage{hyperref}  
 \usepackage{color}
\definecolor{blue}{rgb}{0., 0., 1}
\usepackage{rotating,tabularx}
\hypersetup{colorlinks=true,linkcolor=[rgb]{1.,0.2,0.2},citecolor=[rgb]{0.1,0.4,1.},filecolor=[rgb]{0.7,0.2,0.2},urlcolor=[rgb]{0.7,0.2,0.2}}

\begin{document}
\title{Evidence of extreme ionization conditions and low metallicity in GHZ2/GLASS-z12 from a combined analysis of NIRSpec and MIRI observations}

\correspondingauthor{Antonello Calabro'}
\email{antonello.calabro@inaf.it}

\author[0000-0003-2536-1614]{Antonello Calabr\`o}
\affiliation{INAF - Osservatorio Astronomico di Roma, via di Frascati 33, 00078 Monte Porzio Catone, Italy}

\author[0000-0001-9875-8263]{Marco Castellano}
\affiliation{INAF - Osservatorio Astronomico di Roma, via di Frascati 33, 00078 Monte Porzio Catone, Italy}

\author[0000-0002-7051-1100]{Jorge A. Zavala}
\affiliation{National Astronomical Observatory of Japan, 2-21-1, Osawa, Mitaka, Tokyo, Japan}

\author[0000-0001-8940-6768 ]{Laura Pentericci}
\affiliation{INAF - Osservatorio Astronomico di Roma, via di Frascati 33, 00078 Monte Porzio Catone, Italy}

\author[0000-0002-7959-8783]{Pablo Arrabal Haro}
\affiliation{NSF’s National Optical-Infrared Astronomy Research Laboratory, 950 N. Cherry Ave., Tucson, AZ 85719, USA;}

\author[0000-0002-5268-2221]{Tom J.L.C. Bakx} 
\affiliation{Department of Space, Earth, \& Environment, Chalmers University of Technology, Chalmersplatsen 4 412 96 Gothenburg, Sweden}

\author[0000-0002-4193-2539]{Denis Burgarella}
\affiliation{Aix Marseille Univ, CNRS, CNES, LAM, Marseille, France}

\author[0000-0002-0930-6466]{Caitlin M. Casey}
\affiliation{Department of Astronomy, The University of Texas at Austin, 2515 Speedway Boulevard Stop C1400, Austin, TX 78712, USA}

\author[0000-0001-5414-5131]{Mark Dickinson}
\affiliation{NSF’s NOIRLab, Tucson, AZ 85719, USA}

\author[0000-0001-8519-1130]{Steven L. Finkelstein}
\affiliation{Department of Astronomy, The University of Texas at Austin, 2515 Speedway Boulevard Stop C1400, Austin, TX 78712, USA}

\author[0000-0003-3820-2823]{Adriano Fontana}
\affiliation{INAF - Osservatorio Astronomico di Roma, via di Frascati 33, 00078 Monte Porzio Catone, Italy}

\author[0000-0003-1354-4296]{Mario Llerena}
\affiliation{INAF - Osservatorio Astronomico di Roma, via di Frascati 33, 00078 Monte Porzio Catone, Italy}

\author[0000-0002-9572-7813]{Sara Mascia}
\affiliation{INAF - Osservatorio Astronomico di Roma, via di Frascati 33, 00078 Monte Porzio Catone, Italy}

\author[0000-0001-6870-8900]{Emiliano Merlin}
\affiliation{INAF - Osservatorio Astronomico di Roma, via di Frascati 33, 00078 Monte Porzio Catone, Italy}

\author[0000-0001-7300-9450]{Ikki Mitsuhashi}
\affiliation{National Astronomical Observatory of Japan, 2-21-1, Osawa, Mitaka, Tokyo, Japan}
\affiliation{Department of Astronomy, The University of Tokyo, 7-3-1 Hongo, Bunkyo, Tokyo 113-0033, Japan }

\author[0000-0002-8951-4408]{Lorenzo Napolitano}
\affiliation{INAF - Osservatorio Astronomico di Roma, via di Frascati 33, 00078 Monte Porzio Catone, Italy}
\affiliation{Dipartimento di Fisica, Università di Roma Sapienza, Città Universitaria di Roma - Sapienza, Piazzale Aldo Moro, 2, 00185, Roma, Italy}

\author[0000-0002-7409-8114]{Diego Paris}
\affiliation{INAF - Osservatorio Astronomico di Roma, via di Frascati 33, 00078 Monte Porzio Catone, Italy}

\author[0000-0003-4528-5639]{Pablo G. {P{\'e}rez-Gonz{\'a}lez}}
\affiliation{Centro de Astrobiología (CAB), CSIC-INTA, Ctra. de Ajalvir km 4, Torrejón de Ardoz, E-28850, Madrid, Spain}

\author[0000-0002-4140-1367]{Guido Roberts-Borsani}
\affiliation{Department of Physics and Astronomy, University of California, Los Angeles, 430 Portola Plaza, Los Angeles, CA 90095, USA}

\author[0000-0002-9334-8705]{Paola Santini}
\affiliation{INAF - Osservatorio Astronomico di Roma, via di Frascati 33, 00078 Monte Porzio Catone, Italy}

\author[0000-0002-8460-0390]{Tommaso Treu}
\affiliation{Department of Physics and Astronomy, University of California, Los Angeles, 430 Portola Plaza, Los Angeles, CA 90095, USA}

\author[0000-0002-5057-135X]{Eros Vanzella}
\affiliation{INAF -- OAS, Osservatorio di Astrofisica e Scienza dello Spazio di Bologna, via Gobetti 93/3, I-40129 Bologna, Italy}




\begin{abstract}
GHZ2/GLASS-z12, one of the most distant galaxies found in JWST observations, has been recently observed with both NIRSpec and MIRI spectrographs, establishing a spectroscopic redshift $z_\text{spec}=12.34$, and making it the first system at $z>10$ with complete spectroscopic coverage from rest-frame UV to optical wavelengths. This galaxy is identified as a strong CIV$_{\lambda 1549}$ emitter (EW$=46$ \AA) with many other detected emission lines, such as NIV]$_{\lambda 1488}$, HeII$_{\lambda 1640}$, OIII]$_{\lambda\lambda 1661,1666}$, NIII]$_{\lambda 1750}$, CIII]$_{\lambda\lambda 1907,1909}$, [OII]$_{\lambda\lambda 3726,3729}$, [NeIII]$_{\lambda3869}$, [OIII]$_{\lambda\lambda 4959,5007}$, and H$\alpha$, including a remarkable detection of the OIII Bowen fluorescence line at rest-frame $\lambda =3133$ \AA. We analyze in this paper the joint NIRSpec + MIRI spectral data set. Combining six optical strong-line diagnostics (namely R2, R3, R23, O32, Ne3O2, and Ne3O2Hd), we find extreme ionization conditions, with $\log_{10}$ ($\text{[O {\sc iii}]} _{\lambda\lambda 4959,5007}/\text{[O {\sc ii}]} _{\lambda\lambda 3726,3729}$) $=1.39 \pm 0.19$ and $\log_{10}$ ($\text{[Ne {\sc iii}]} _{\lambda 3869}/\text{[O {\sc ii}]} _{\lambda\lambda 3726, 3729}$) $=0.37 \pm 0.18$ in stark excess compared to typical values in the ISM at lower redshifts. 
These line properties are compatible either with an AGN or with a compact, very dense star-forming environment (\sigmaSFR $\simeq 10^2$-$10^3$ \msun/yr/kpc$^2$ and \sigmaMass $\simeq 10^4$-$10^5$ \msun/pc$^2$), with a high ionization parameter ($\log_{10}$(U) $=-1.75 \pm 0.16$), a high ionizing photon production efficiency $\log(\xi_{\rm ion}) = 25.7_{-0.1}^{+0.3}$, and a low gas-phase metallicity (also confirmed by the direct, T$_e$ method) ranging between $4\%$ and $11\%$ Z$_\odot$, indicating a rapid chemical enrichment of the ISM in the last few Myrs. 
These properties also suggest that a substantial amount of ionizing photons ($\sim 10\%$) are leaking outside of GHZ2 and starting to reionize the surrounding intergalactic medium, possibly due to strong radiation driven winds. The general lessons learned from GHZ2 are the following: (i) the UV to optical combined nebular indicators are broadly in agreement with UV-only or optical-only indicators. (ii) UV+optical diagnostics fail to discriminate between an AGN and star-formation in a low metallicity, high density, and extreme ionization environment. (iii) comparing the nebular line ratios with local analogs may be approaching its limits at $z \gtrsim 10$, as this approach is potentially challenged by the unique conditions of star formation experienced by galaxies at these extreme redshifts. 
\end{abstract}



\section{Introduction}\label{sec:intro}

The exciting discovery of galaxies in the early Universe was among the science drivers and the most eagerly expected results. The expectations have been surpassed by the earliest results from JWST, with multiple NIRCam imaging surveys \citep{castellano22,castellano23a,naidu22,finkelstein22,harikane22,donnan23a,donnan23b,franco23,casey23} revealing dozens of galaxy candidates at redshifts beyond $\sim 10$, when the Universe was less than $\sim 500$ Myr old. Their measured number density and luminosity are in stark excess compared to those expected from virtually any pre-JWST theoretical or empirical model \citep{finkelstein23a,arrabal23a,arrabal23b_nature,bouwens23}. 

This discovery calls for significant changes to our understanding of the physics of early galaxies. Multiple physical scenarios have been proposed to explain the excess of bright galaxies, including a higher stellar-to-halo mass ratio and star-formation efficiency at early epochs \citep{harikane23,mason23,harikane24}, an increased weight of Pop III and metal poor stars with a top-heavy IMF \citep{trinca23}, negligible UV dust optical depth due to radiation-driven outflows at super-Eddington luminosities \citep{ferrara23a}, and contribution of early AGN activity, primordial black holes or other exotic particles to the UV photon budget \citep{cappelluti22}. To make progress in this field, and also assess the role of the high-$z$ galaxies population in cosmic reionization, we must move from simple detection to the physical characterization by probing their rest-frame UV and optical emission lines with spectroscopic observations. 

The lack of Balmer and metal lines at $\lambda > 4000$ \AA, which all fall beyond the spectral coverage of NIRSpec (the most sensitive spectrograph on board the JWST), have hampered so far a thoughout characterisation of the most distant sources.
For this reason, a first simple strategy to approach the primordial Universe has been that of probing the average conditions of galaxies at increasing redshifts, and then try to extrapolate those at $z>10$. 
Indeed, NIRSpec has already opened a new frontier by probing with unprecedented depth the ionization and metallicity of statistical samples of galaxies down to low mass ($\sim 10^7$ \msun) from $z\simeq4$ to $z\simeq10$. Several studies \citep[e.g.,][]{trump23,curti23a,curti23b,nakajima23} have found that typical galaxies at the Epoch of Reionization (EoR) have relatively low Z$_\text{gas}$ values, but not significantly different from those observed at $z \gtrsim 2$, remaining close or above $10\%$ Z$_\odot$.  
If these trends persist beyond $z \simeq 10$, it will be difficult to explain the excess of UV luminosity density in the pre-EoR as due to extremely metal poor stellar populations. 
On the other hand, several studies have found a significant evolution in ionizing strength from cosmic noon to $z\sim9$ \citep{pahl20,reddy23,trump23,cameron23}. 
This leaves us with a question about how cosmic ionization evolves at earlier times, and what are the properties of galaxies that may have started the reionization in the first 500 Myr of our Universe. 
The next unavoidable step is to directly observe spectroscopically this unexpectedly UV-bright population at $z>10$. 
A handful of galaxies have been recently observed (with NIRSpec) in the redshift range from $10$ to $12$, such as Maisie's Galaxy and CEERS2$\_$588 \citep{arrabal23b_nature}, GN-z11 \citep{bunker23}, and MACS0647-JD \citep{hsiao23}. However, longer wavelength observations with MIRI are required to obtain a complete rest-frame optical spectral coverage. 

The combination of NIRSpec and MIRI spectroscopy was recently achieved for 
GHZ2/GLASS-z12 (hereafter simply GHZ2), which is another remarkable example and one of the most representative of the UV-bright galaxy population at $z>10$. Lying in a small area of $\simeq 10$ arcmin$^2$ in the background of the galaxy cluster Abell 2744, it was initially identified in imaging by \citet{castellano22} and \citet{naidu22} within the GLASS-JWST Early Release Science program \citep{treu22}. Independent teams have estimated its photometric redshift $z_{phot}$ in a range between $11.9$ and $12.4$, using different methods \citep{naidu22,castellano22,harikane22,donnan23b,atek23}. According to the SED fitting estimates, GHZ2 has among the highest UV luminosity at such high redshift (M$_\text{UV}=-20.5$ mag), and it may have already built $\geq10^9$ \msun in stars, a factor of $\sim 3$ higher than expected from the maximum SHMR at this redshift \citep{behroozi20}. 
 
This galaxy was observed with the JWST Near Infrared Spectrograph (NIRSpec) in low-resolution prism configuration on Oct. 24th 2023 by the Program GO-3073 (PI M. Castellano), which has revealed the detection of multiple emission lines at $z=12.34$ including NIV]$_{\lambda 1488}$, HeII$_{\lambda 1640}$, OIII]$_{\lambda\lambda 1661,1666}$, NIII]$_{\lambda 1750}$, CIII]$_{\lambda\lambda 1907,1909}$, OIII$_{\lambda 3133}$, [OII]$_{\lambda\lambda 3726,3729}$, [NeIII]$_{\lambda3869}$, and a bright CIV$_{\lambda 1549}$ with an equivalent width EW $=46$ \AA, placing this source in the category of strong CIV emitters \citep{castellano24}. 
Despite the multiple lines detected, the UV spectral properties of GHZ2 are unconclusive on the star-forming or AGN nature of the source. 
Even the surprising detection of the OIII Bowen fluorescence line at $3133$ \AA\ is not a decisive evidence of AGN as its emission could be also associated to X-ray binaries and planetary nebulae \citep{pereira99,selvelli07,liu93}. In addition, the absence of the high-ionization NV$_{\lambda 1240}$ and [NeV]$_{\lambda 3426}$ lines, and the low upper limit on the [NeIV]$_{\lambda 2424}$/NIV]$_{\lambda 1488}$ ratio are likely inconsistent with the AGN hypothesis and more in line with a star-forming scenario, according to the models of \citet{feltre16} and \citet{gutkin16} \footnote{We note that those models assume subsolar to solar N/O abundance ratios that might differ from those in GHZ2 and other nitrogen enriched galaxies.}. 
To probe the rest-frame optical lines, GHZ2 was also targeted with the JWST Mid-Infrared Instrument (MIRI) Low Resolution Spectrometer (LRS), and observed in the same period (25-29 October 2023) by the program GO-3703 (PI J. Zavala).
As described in \citet{zavala24}, we detect H$\alpha$ and [OIII]$_{\lambda\lambda 4959,5007}$. 
No broad H$\alpha$ components were detected, limited by the low spectral resolution,  
leaving still open the interpretation on the nature of this object.  

These observations make GHZ2 the most distant galaxy for which we have spectroscopic measurements covering the full UV to optical spectral range. 
The purpose of this paper is to combine the information from the emission lines detected in NIRSpec and MIRI, to provide a comprehensive understanding and a more robust assessment of its ionization and metallicity properties. 
In particular, we will analyze in a systematic way the following six, most widely adopted, strong line diagnostics (see \Citealt{kewley08} for a thorough discussion) : 
R2 = $\log (\text{[O {\sc ii}]} _{\lambda\lambda 3726,3729} / \text{H} \beta) $, 
R3 = $\log (\text{[O {\sc iii}]} _{\lambda 5007} / \text{H} \beta) $, 
R23 = $\log \left( (\text{[O {\sc iii}]} _{\lambda\lambda 4959,5007} + \text{[O {\sc ii}]} _{\lambda\lambda 3726,3729} ) / \text{H} \beta \right)$ , 
O32 = $\log (\text{[O {\sc iii}]} _{\lambda\lambda 4959,5007} / \text{[O {\sc ii}]} _{\lambda\lambda 3726,3729} ) $, 
Ne3O2 = $\log (\text{[Ne {\sc iii}]} _{\lambda 3869} / \text{[O {\sc ii}]} _{\lambda\lambda 3726, 3729} )$,
and Ne3O2Hd = $\log \left( ( \text{[Ne {\sc iii}]} _{\lambda 3869} + \text{[O {\sc ii}]} _{\lambda\lambda 3726, 3729} ) / \text{H} \delta \right) $.
With this line dataset, we will analyze the metallicity and ionization with calibrations based on local analogs of high redshift systems, and we will further investigate the star-forming or AGN nature of the source.
Additionally, this will enable us to conduct comparisons with data inferred solely from NIRSpec or MIRI, or from fitting NIRCam photometry with stellar population models. Such analyses will offer insights into the optimal strategies for future spectroscopic follow-ups of the earliest galaxies. The spectro-photometric analysis combining NIRCam + NIRSpec + MIRI will be presented in a different paper. 

The paper is organized as follows. In Section \ref{sec:methods}, we briefly summarize the NIRSpec and MIRI observations of GHZ2, the derivation of the emission line fluxes, and the physical properties obtained independently in the two studies. In Section \ref{sec:results}, we combine the emission lines detected in the two spectra to better assess the ionization and metallicity of GHZ2 via line ratios, and we discuss the possibility of distinguishing between AGN or stellar photoionization. 
We present our conclusions in Section \ref{sec:conclusions}.

Throughout the paper we adopt AB magnitudes \citep{oke83}, a \citet{chabrier03} initial mass function (IMF), a solar metallicity of 12 + log(O/H) = 8.69 \citep{asplund09},  and a flat $\Lambda$CDM concordance model (H$_0$ = 70.0~km~s$^{-1}$~Mpc$^{-1}$, $\Omega_M=0.30$). 

\section{Methodology}\label{sec:methods}

\subsection{Observations and Data Reduction}\label{sec:obs}

While the full description of the observations and spectral reduction is presented in \citet{castellano24} and \citet{zavala24} (hereafter cited as simply C24 and Z24), we highlight here the most important features. 
GHZ2 was observed with NIRSpec in PRISM-CLEAR configuration (i.e., with spectral resolution R ranging from $30$ at $\lambda=0.6\ \mu m$ to $330$ at $\lambda=5.3\ \mu m$), adopting 3-shutter `slits' with a three-point nodding for optimal background subtraction, reaching a total exposure time of $19701$s in three separate visits. 
The data were reduced with the standard calibration pipeline provided by STScI \citep[version 1.13.4,][]{bushouse24} following the methodology of \citet{arrabal23a}, which provides the full wavelength and flux calibrated 2D and 1D spectrum. The output NIRSpec spectrum was also corrected for residual slit and aperture losses by matching the detected continuum level with the latest available broadband NIRCam photometry (Merlin et al. in prep.), as already done in \citet{napolitano24}. 

The MIRI observations were conducted with the low resolution spectrometer (LRS) slit mode, which provides a resolving power of R$\simeq50$ to $200$ over $5$-$12 \mu m$, with an `along-slit' dithering mode and a total integration time on-source of $9$h. The spectral reduction was performed using the same version of the STScI pipeline adopted for NIRSpec, with background subtraction applied on each dither position, and final extraction of the 2D and 1D spectrum performed as described in Z24.  

\subsection{Main physical properties of GHZ2}\label{sec:physicalproperties}

Multiple emission lines were detected through the NIRSpec and MIRI observations, as explained in Section \ref{sec:intro}, which have allowed us to constrain the spectroscopic redshift $z_\text{spec}$ of the galaxy. We adopt here the z$_\text{spec}=12.342 \pm 0.009$ derived from a weighted average of the measurements of the best resolved, high-SNR lines in the NIRSpec spectrum (C24), and in good agreement with the MIRI-derived estimate of $12.33\pm0.02$ (Z24). This is also in agreement with that estimated from the high-resolution spectrum obtained with the VLT X-SHOOTER instrument by the program 110.244H.001 (PI E. Vanzella), in which CIV$_{\lambda 1548}$ was detected at a S/N $=4.4$. 
We remark that for investigating the nebular properties in the combined NIRSpec + MIRI spectrum, we consider throughout this work the line fluxes and uncertainties reported in the two companion papers. 

We adopt a stellar mass \mstar\ of $\log_{10}$ \mstar/\msun $=8.91_{-0.28}^{+0.13}$, derived by Z24 through fitting the NIRCam photometry and the emission lines with the SYNTHESIZER-AGN code \citet{perezgonzalez03}, with \citet{bruzual03} stellar populations models, assuming a Chabrier initial mass function with stellar mass limits between $0.1$ and $100$ \msun, a double burst with delayed-exponential law, and nebular emission grids computed with CLOUDY v23 \citep{chatzikos23}. This stellar mass is consistent with the estimate of $\log M_{\star} / M_{\odot} = 9.05^{+0.10}_{-0.25}$ derived, 
independently, by C24 with BAGPIPES \citep{carnall18} using only the photometric data, assuming the Binary Population and Spectral Synthesis (BPASS) stellar population models v2.2.1 \citep{eldridge17,stanway18}, and a double power law star-formation history (SFH). 
We also adopt the best-fit mass-weighted age of $28_{-14}^{+10}$ Myr from the first study, which implies that $\simeq 60\%$ of the total \mstar was formed in the last $30$ Myr (Z24). 

We adopt for GHZ2 the SFR obtained by Z24 from the \Ha line in the MIRI spectrum, which yields SFR $=9 \pm 3$ \msun/yr. This assumes the calibration of \citet{reddy22} as SFR $=$ L$_{H\alpha}$ $\times$ $10^{-41.67}$, which is the most suited for the subsolar metallicities expected in the early Universe, reflecting the greater efficiency of ionizing photon production in metal poor stellar populations, and harder ionizing spectra due to binary star interactions. 
From the SED fitting presented above, Z24 derive a SFR that is consistent with the \Ha based value, and a very low dust attenuation in V band (A$_V = 0.1_{-0.1}^{+0.2}$, which adds to the very blue UV slope measured for the galaxy ($\beta=-2.39 \pm 0.07$) by C24. 
Assuming negligible dust attenuation and case B recombination, Z24 rescale the H$\alpha$ flux to estimate an intrinsic \Hb and H$\delta$ flux, which are consistent with the observed upper limits of the two undetected lines. 

Several studies have found that low mass, extreme emission line galaxies at high redshift may have balmer decrements significantly smaller than case B recombination by factors of up to $\sim 10\%$ \citep{stiavelli23,topping24,yanagisawa24}. While the physical origin of these balmer decrement anomalies is still unclear and possibly due, among all, to density bounded geometries or optically thick, excited neutral gas clouds \citep[see ][]{yanagisawa24,scarlata24,mcclymont24}, we note that, assuming the most conservative \Ha/\Hb\ observed ratio of $2.55$ \citep{topping24,yanagisawa24}, would produce small variations ($\sim 6\%$) of balmer based line indices, and $\sim 10\%$ lower gas-phase metallicities. Therefore, this would not significantly affect our results and would not alter the main conclusions of this paper.
We remark that additional SED fitting methods were tested in Z24, which yield lower stellar masses (by $\sim 0.6$ dex) and dust attenuations A$_V$ ranging $0$-$0.3$ mag. Following these results, we incorporate an uncertainty of $+0.3$ mag on A$_V$ in the following analysis. Alternative and more sophisticated fitting procedures will be investigated in future works. 

We assume an effective radius of $r_e = 105 \pm 9$ pc measured by \citet{yang22}, from which we calculate the SFR surface density \sigmaSFR as SFR/($2 \times \pi \times r_e^2$), yielding $\log$ \sigmaSFR/(\msun yr$^{-1}$ kpc$^{-2}$) $=2.11_{-0.25}^{+0.35}$. We note that a smaller radius ($=34 \pm 9$~pc) was measured by \citet{ono23} \footnote{We note that $r_e$ refers to the semi-major axis in \citet{yang22}, while it is defined as the circularized radius in \citet{ono23}. For sanity check, we have independently fitted a sersic profile to GHZ2 with the code galight \citep{ding21}, finding that this galaxy is unresolved, with an upper limit on $r_e$ (corrected for lensing) of $80$ pc, thus more consistent with \citet{ono23}. To be conservative, we consider both previous measurements in our analysis. In any case, our final conclusions are not affected by the choice of $r_e$.}, in which case $\log$ \sigmaSFR/(\msun yr$^{-1}$ kpc$^{-2}$) would rise to $3.1 \pm 0.4$.  
We finally remark that all the physical properties introduced above have been properly corrected to account for gravitational lensing, using the moderate magnification ($\mu=1.3$) estimated by \citet{bergamini23}.

\subsection{Comparison to photoionization models}\label{sec:photoionization_models}

To understand the nebular and ionizing source properties of GHZ2, it is useful to compare our observed emission line ratios to those predicted by photoionization models. 
To this aim, we consider the line predictions derived with the photoionization models described in \citet{calabro23}, which are computed with the python package pyCloudy v.0.9.11 \footnote{\url{https://github.com/Morisset/pyCloudy/tree/0.9.11} }, 
running with version 17.01 of the Cloudy code \citep{ferland17}. 

In brief, star-forming galaxies are modeled with a spherically symmetric, radiation-bounded, shell of gas surrounding a population of young (O and B type) stars, with the incident radiation field derived from BPASS stellar population models \citep{eldridge17} with an IMF extending to $100$ \msun and continuous star-formation in the past $30$ Myrs, to match the measured average age of GHZ2 (Z24). 
We also analyze the emission predicted by AGN models, in which the continuum is built using the default 'AGN' prescription in Cloudy, with a multiple power law continuum assuming a 'blue bump' temperature of $10^6$ K, spectral energy indices of $\alpha_{UV} = -0.5$, $\alpha_x = -1.35$, $\alpha_{ox}=-1.4$ \citep{groves04}, respectively in UV, X-rays, and from the optical to the X-ray range. We consider the metallicity range from $0.05$ and $1$ times solar (i.e., $0.05$, $0.1$, $0.15$, $0.2$, $0.3$, $0.4$, $0.5$, $0.7$, $1$) for SF galaxies, and from $0.05$ and $2$ times solar (i.e., $0.05$, $0.1$, $0.2$, $0.3$, $0.4$, $0.5$, $0.7$, $1$, $2$) for AGNs, with the solar reference consistent with our definition in Section \ref{sec:intro}. In all cases, we derive predictions for four different ionization parameters $\log$(U)$=-3$, $-2.5$, $-2$, and $-1.5$, and for three gas density values ($10^2$, $10^3$, and $10^4$ $cm^{-3}$). 
Regarding dust depletion, the metals are depleted in the beginning of our CLOUDY calculations, and we consider that this depletion is metallicity dependent, as discussed in \citet{calabro23}. In particular, for the elements analyzed in this work, Ne is a refractory element in all conditions \citep{gutkin16}, while the depletion factor of oxygen (0.2 dex at solar metallicity) is expected to decrease significantly when going to sub-solar metallicity, especially at $\lesssim 0.1$ Z$_\odot$ \citep{vladilo11,decia16}, hence we assume that it is negligible for GHZ2. This is also reasonable considering the very low dust attenuation (hence dust content) that we infer for this galaxy.

We also explore the density-bounded scenario for the nebula. We have modeled this case by varying the stopping criterion in Cloudy from a lyman continuum (LyC) optical depth $= 10$ (fully ionization bounded case) to $0.1$ (fully density bounded case). This would correspond to an escape fraction of ionizing photons going from $0\%$ to $100\%$, according to \citet{plat19}. 
As noted by the other studies \citep[e.g.,][]{pellegrini12,jaskot13,nakajima20}, a density bounded scenario tends to increase the flux of high ionization species with respect to low ionization species. In our case, it would increase the O32 and Ne3O2 indices, mimicking the effect of a high ionization parameter.
However, we find that, with the modest escape fraction ($\sim 10\%$) estimated for GHZ2 (see later in Section 4), the increase in O32 and Ne3O2 would be lower than $0.1$ dex, thus would not change the main interpretation of this paper. 
We finally note that, following a similar argument to \citet{reddy23}, the high electron density ($\geq 10^3$ cm$^{-3}$, Z24) and the high SFR of this galaxy suggest that we are closer to a radiation bounded geometry.  

We refer to \citet{calabro23} for a more detailed description of the models, including element abundances and dust depletion. 
Considering different AGN models among those tested in \citet{calabro23}, or different setups for the star-forming models, does not significantly affect the predicted line ratios and does not alter the conclusions of this paper.

\section{Results and discussion}\label{sec:results}

\subsection{Combining NIRSpec and MIRI emission line ratios}\label{line_ratios}

\bigskip
We explore in this first subsection the ionizing and excitation conditions of the ISM, while in the following one we will investigate whether the ionization field is of stellar or AGN origin. The measured values of the various line ratio indices used in this paper are reported in Table~\ref{tab:metallicity}.

\subsubsection{Excitation vs ionization diagnostics}\label{sec:R23_O32}


\begin{figure*}[t!]
\centering
\vspace{-0.14cm}
\includegraphics[angle=0,width=0.90\linewidth,trim={0cm 0cm 0.cm 0cm},clip]{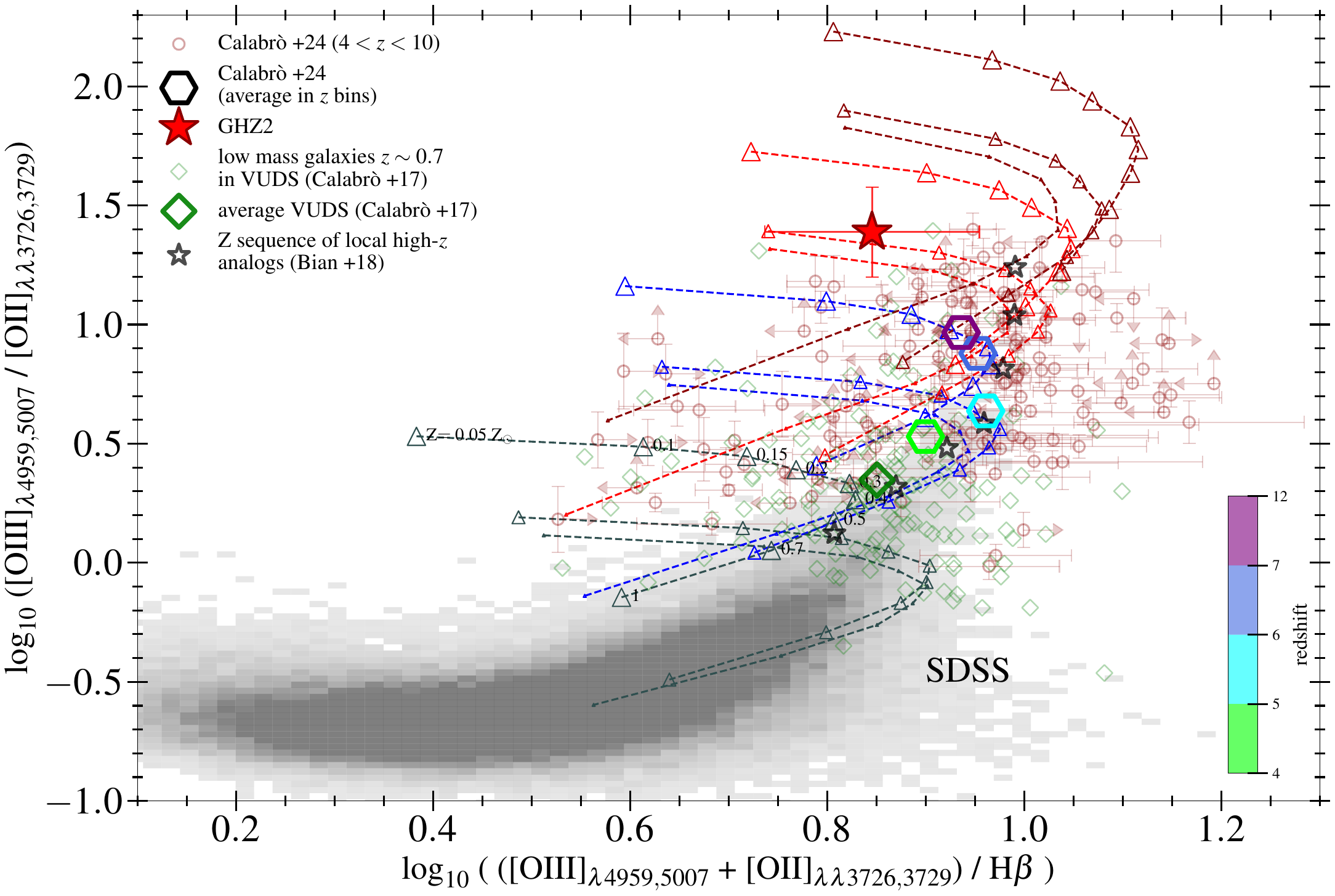}
\includegraphics[angle=0,width=0.90\linewidth,trim={0cm 0cm 0.cm 0cm},clip]{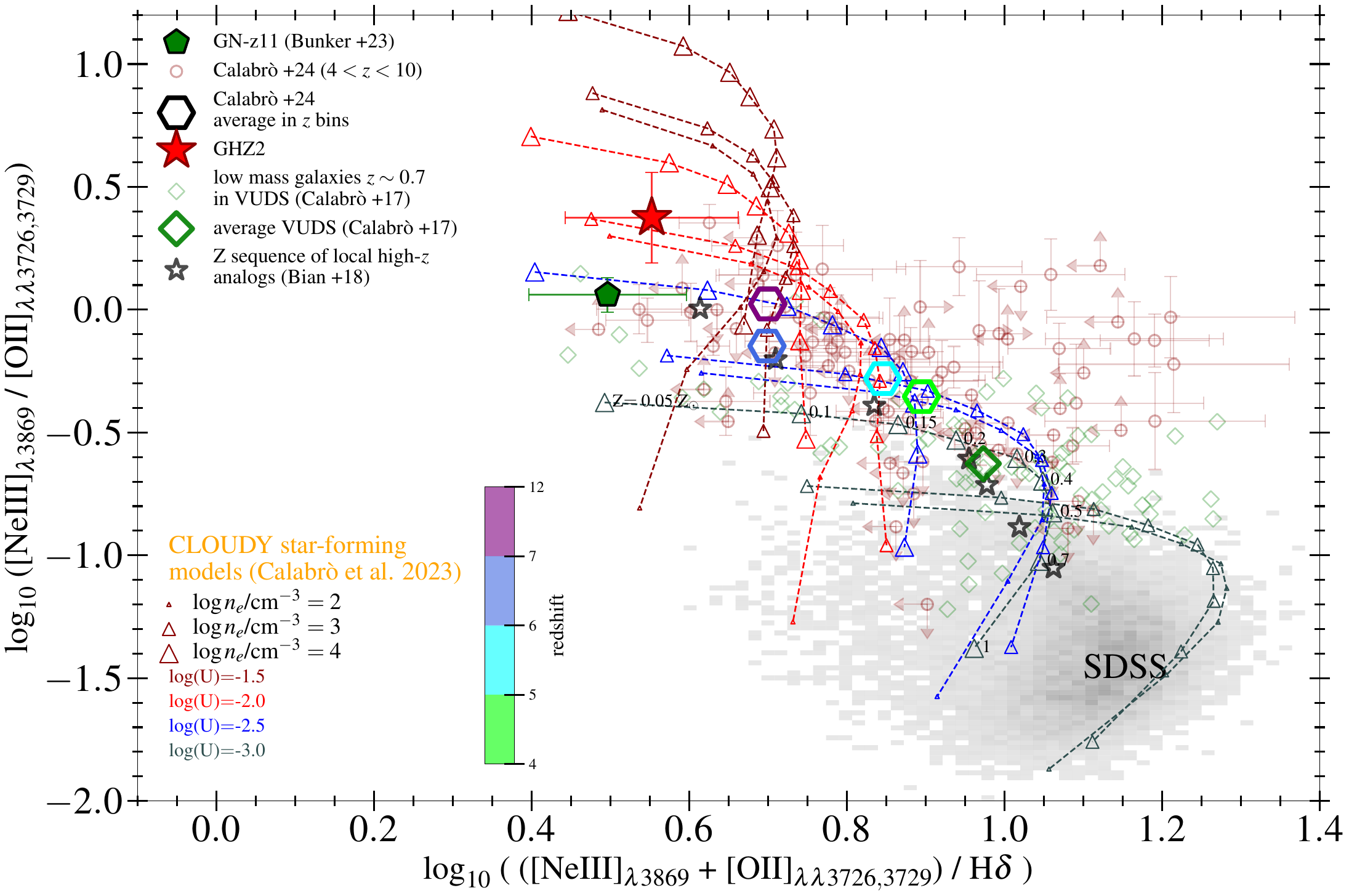}
\vspace{-0.16cm}
\caption{\textit{Top:} The R23 vs O32 diagram. \textit{Bottom:} The Ne3O2Hd vs Ne3O2 diagram. GHZ2 (big red star), is compared to the NIRSpec sample of SF galaxies from \citet{calabro24}. Local SDSS SF galaxies are shown with a binned 2D gray histogram, while $z\sim0.7$ low-mass star-forming galaxies from \citet{calabro17} are shown with green empty diamonds. Overplotted are SF model predictions computed with pyCLOUDY from \citet{calabro23}, assuming the following ranges of $\log(U)$ ($-3$, $-2.5$, $-2$, $-1.5$), Z ($0.05$, $0.1$, $0.15$, $0.2$, $0.3$, $0.4$, $0.5$, $0.7$, and $1$ $\times$ Z$_\odot$), and $\log$ $n_e/cm^{-3}$ ($2$, $3$, $4$), as shown in the legend in the bottom panel. Both diagrams suggest a $\log$(U) $\simeq -2$ and a low metallicity between $0.05$ and $0.1 \times$ Z$_\odot$. 
} 
\label{fig:excitation_ionization}
\end{figure*} 

One of the most valuable nebular diagnostics is built by comparing the R23 and the O32 line indices, which probe respectively the excitation and the ionization state of the gas. This diagram has been widely used to study the ISM conditions both in the local Universe and at higher redshifts \citep[e.g.,][]{maiolino08,sanders16,onodera16,nakajima20,flury22b_tables,schaerer22b_JWST,reddy23}. 
To put in context our results, we consider the NIRSpec sample recently observed by the CEERS and GLASS JWST spectroscopic surveys \citep{mascia23,calabro24}, representative of the star-forming galaxy population at the EoR in the mass range $7 < \log$(\mstar/\msun) $<10.5$, and a sample of similarly low-mass star-forming galaxies (\mstar down to $10^{7}$ \msun) at intermediate redshifts ($z \sim 0.7$) from the VUDS survey \citep{calabro17}. 
Finally, we explore the parameter space occupied by galaxies from the Sloan Digital Sky Survey (SDSS) as a benchmark for typical star-forming systems in the local Universe \citep{kauffmann03}. We also consider a subset of metal poor and high ionization systems selected in the local Universe to mimick the properties of high redshift galaxies \citep{bian18}. 

We show the R23-O32 diagram in Fig. \ref{fig:excitation_ionization}-\textit{top}. We can notice an overall increase of the ISM ionization when we go from the local Universe, probed by the SDSS, to intermediate redshifts ($z \sim 0.7$), probed by the VUDS survey, and then to galaxies at $z\geq4$, probed by the NIRSpec galaxies. 
GHZ2 lies in the upper part of the diagram, with an O32 index of $1.39 \pm 0.19$ and an R23 index of $0.88 \pm 0.12$. While the latter value is similar to what is found also in lower redshift systems, the O32 is higher than in typical star-forming galaxies at the EoR, and also compared to the median properties of the so-called local analogs of high redshift galaxies \citep{bian18}, highlighted in Fig. \ref{fig:excitation_ionization} with empty black stars for bins of decreasing metallicity. 
GHZ2 has also a similar O32 to that measured for the strongly lensed galaxy MACS1149-JD1 (at $z=9.1$) identified by \citet{stiavelli23}. 

Comparing the position of GHZ2 in this diagram to photoionization model predictions for star-forming galaxies, we find that it lies in the low metallicity branch (Z $< 0.3$ Z$_\odot$), where R23 turns over and starts to decline in more metal poor systems, and its ISM is consistent with a high ionization parameter ($\log$(U) $\sim -2$) and a high electron density $n_e$ of the gas between $10^{3}$ cm$^{-3}$ and $10^{4}$ cm$^{-3}$, or, at a fixed $n_e$ of $10^{3}$ cm$^{-3}$, with an even higher ionization ranging $-2 \leq $ $\log$(U) $\leq -1.5$.
We note that a similar range of $n_e$  ($\geq 10^{3}$ cm$^{-3}$) is suggested by ALMA observations \citep{popping23,zavala24} by the non detection of the \OIII $_{\lambda 88 \mu m}$ line (assuming an electron temperature T$_e$ $=10^4$ K). The inferred range of $\log$(U) would be in agreement with the estimates of $-1.78\pm0.28$ and $-1.4_{-0.3}^{+0.2}$ based on two SED fitting approaches by C24 and Z24. 

We calculate directly the ionization parameter from the O32 index using an extrapolation toward higher O32 of the empirical relation derived by \citet{papovich22} with galaxies at redshifts $1.1 < z < 2.3$ observed by the CLEAR survey. This yields $\log_{10}$(U) $=-1.75 \pm 0.16$, which is within the range suggested by the previous models comparison and by the SED fitting method. This again points towards very strong ionizing conditions that are typically not seen in low redshift galaxies.

We now compare the results of the R23 - O32 diagran with those from another `excitation vs ionization' diagram, involving the $\log$ \NeIII $_{\lambda3869}$ + \OII$_{\lambda\lambda 3726,3729}$)/H$\delta$\ (Ne3O2Hd) index and Ne3O2. This diagram can be used 
for galaxies at $9 \lesssim z \lesssim 12$, in which the \Hb and \xOIII\ lines fall outside of the NIRSpec coverage (see recent applications in \Citealt{bunker23} for GN-z11, and in \Citealt{robertsborsani24}).  
H$\delta$ is typically fainter than \NeIII\ or \OII, but its flux can be rescaled from that of Balmer lines at longer wavelengths after accounting for dust attenuation.

For GHZ2, we follow the example of Z24, and derive H$\delta$ from \Ha assuming case-B recombination \citep{osterbrock06} and negligible dust attenuation, which is supported by the blue UV continuum slope (C24) and the ALMA detection limits \citep{bakx23}.
This yields a Ne3O2Hd index of $0.59 \pm 0.12$. We also measure for this galaxy a Ne3O2 value of $0.37 \pm 0.18$. A similar procedure was used to derive Ne3O2Hd in lower redshift comparison samples.

The position of GHZ2 in the Ne3O2Hd - Ne3O2 parameter space is shown in Fig. \ref{fig:excitation_ionization}-\textit{bottom}. GHZ2 is in the upper left envelope of the distribution that extends from the local SDSS and the intermediate-$z$ VUDS galaxies in the lower right corner of the diagram to the top left part occupied by the NIRSpec galaxies at $z>4$, following a sequence of increasing Ne3O3 (and decreasing Ne3O2Hd) as a function of redshift. We highlight that GHZ2 has even more extreme properties than typical star-forming galaxies at the EoR. Its position is instead similar to that of GN-z11 \citep{bunker23} and MACS1149-JD1 \citep{stiavelli23}, in which they detect H$\delta$, even though GHZ2 has a more extreme ionization traced by Ne3O2. 
The EW and line ratios of GHZ2 are also similar to those of the compact galaxy \rxc\ at $z=6.1$ \citep{topping24}, as noted by C24. In the analyzed diagrams, this would be the only object more extreme than GHZ2, as having an O32 $\simeq 2.3$ and Ne3O2 $\simeq 1.2$. 
The Ne3O2 index measured for GHZ2 also exceeds the parameter range covered by the sample of \citet{bian18}, indicating that local analogs of high redshift galaxies do not have the same extreme conditions found in this galaxy. 
Similarly to the R23 - O32 diagram, also the Ne3O2Hd - Ne3O2 one favors high electron densities $\geq 10^3$ cm$^{-3}$, low metallicity ($0.05 <$ Z/Z$_\odot$ $<0.1$), and high ionization parameter $\log$(U) $\sim -2$.

Overall, this diagram provides a picture that is consistent with that obtained with the classic R23 vs O32 diagram, indicating the presence of a hard ionizing radiation field in GHZ2, and more extreme ionization properties than typically found in star-forming galaxies.  

\subsubsection{Investigating the star-forming or AGN nature with the OHNO diagram}\label{sec:OHNO}

\begin{figure*}[t!]
\centering
\includegraphics[angle=0,width=0.97\linewidth,trim={0cm 0cm 0.cm 0cm},clip]{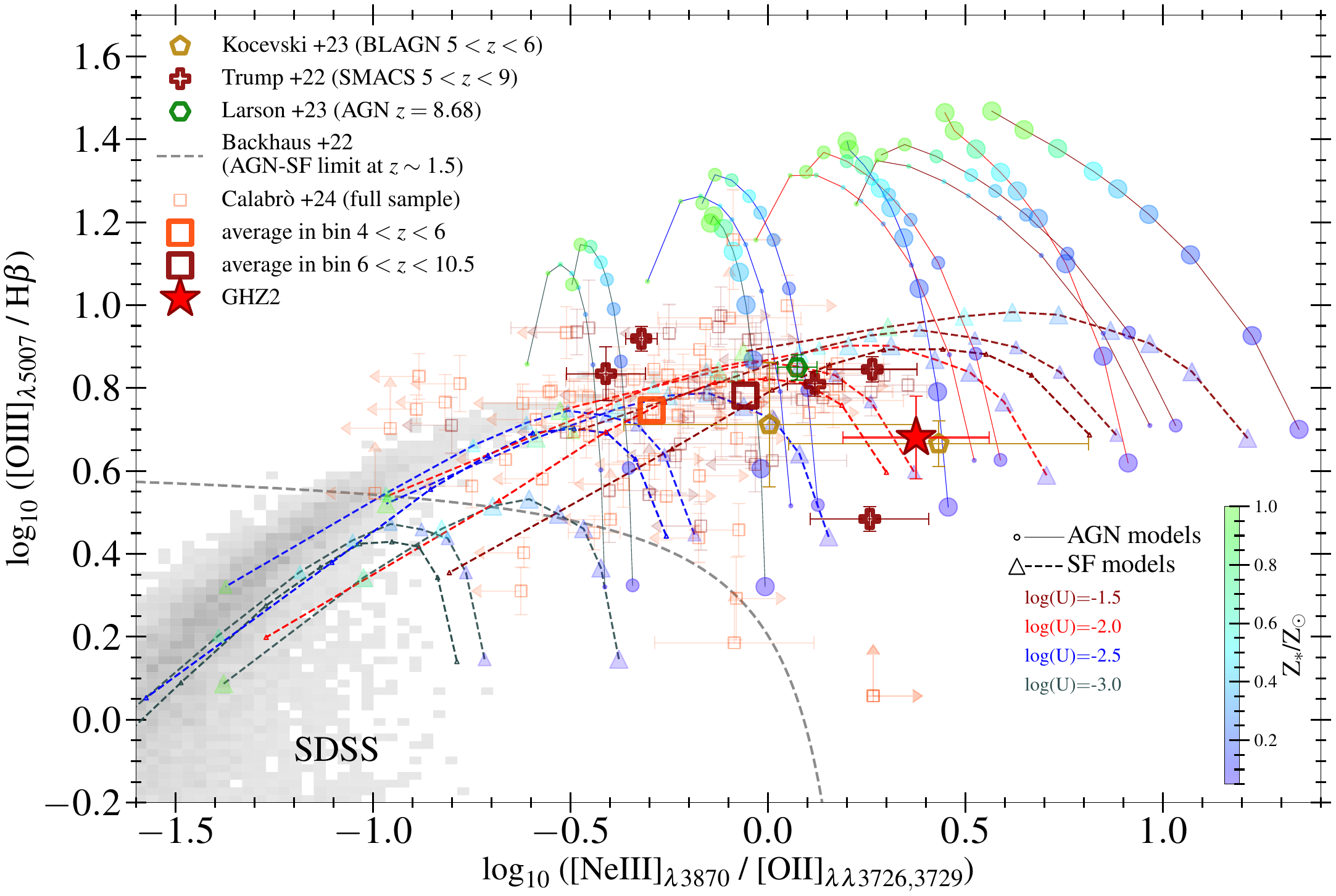}
\vspace{-0.01cm}
\caption{The $\log$ \OIII$_{\lambda\lambda5007)}$/\Hb\ (R3) - $\log$ \NeIII\ $_{\lambda3869}$/\OII$_{\lambda\lambda3726,3729}$ (Ne3O2) diagram (`OHNO'). The position of GHZ2 is shown with a big red star, while other observations of star-forming galaxies and broad line AGNs identified at $z>5$ by recent JWST observations are included as big colored polygons. The predictions of photoionization models from \citet{calabro23} are overplotted for comparison. The models are shown for three representative electron densities $n_e$ $=10^2$, $10^3$, and $10^4$ cm$^{-3}$ (increasing size symbols), and cover a variety of metallicities (from $0.05$ to solar, following the colorbar in the bottom right corner) and ionization parameters ($-3$,$-2.5$, $-2$, $-1.5$, following the convention of Fig. \ref{fig:excitation_ionization}). 
}
\label{fig:OHNO}
\end{figure*} 

The O32 and Ne3O2 index values are a clear sign of a hard ionizing source powering GHZ2 (see also C24 and Z24). Understanding whether this hard ionization field is due to star formation or an AGN is fundamental to reach a physical interpretation on the nature of massive and blue galaxies in the bright end of the luminosity function at $z>10$.  
We have shown in the two companion papers that it is not possible to distinguish between AGN and massive young stellar population photoionization on the basis of the NIRSpec-PRISM and MIRI-LRS spectrum alone, lacking enough sensitivity and spectral resolution.  

The combination of NIRSpec and MIRI allows us to test the nature of GHZ2 with another diagram by comparing the R3 and the Ne3O2 line indices, dubbed the OHNO diagram. 
This diagnostic has been successfully applied to classify galaxies from the local universe to cosmic noon \citep{zeimann15,backhaus22,cleri23}. At redshift $\sim1.5$, a separation criterion was also proposed by Backhaus et al. 2022, according to which AGNs have significantly higher R3 and Ne3O2 than star-forming galaxies.
These studies have found that from $z=1$ to $z=10$, Ne3O2 increases on average, while R3 remains relatively constant. This can be attributed to higher gas densities and ionization parameters in HII regions at higher redshifts, and perhaps also to a greater contribution of AGNs to photoionization.

For GHZ2 we measure an R3 index of $0.72 \pm 0.11$. The resulting position in the OHNO diagram is shown in Fig. \ref{fig:OHNO}, compared to photoionization models predictions. Other high redshift sources recently observed by JWST in the epoch of reionization are also included \citep{trump23,kocevski23,larson23,calabro24}. First, we can see that the source lies in the AGN region according to the separation criteria derived at $z\sim1.5$ by \citet{backhaus22}, and is consistent with the prediction of AGN models with subsolar metallicities ($0.05 <$  Z$_{gas}$/Z$_\odot$ $< 0.1$), high ionization parameters ($\log_{10}$U $\sim -2$), and electron densities $n_e$ ranging $10^{2}$-$10^{3}$ cm$^{-3}$. Another possible interpretation would be a higher $n_e \sim 10^{4}$ cm$^{-3}$ and a slighly lower $\log_{10}$U $\sim -2.5$. Its position closely resembles that of confirmed broad-line AGNs at $z>5$ from \citet{kocevski23} and \citet{larson23}. 
However, the galaxy is also consistent with star-forming models, overlapping with the parameter space expected for HII regions with high electron density ($n_e \geq 10^3$ cm$^{-3}$), high ionization parameter ($\log_{10}$U $\sim -2$), and low metallicity ($0.05 <$  Z$_{gas}$/Z$_\odot$ $< 0.1$), in remarkable agreement with the results inferred from the two star-forming diagnostics analysed above (Fig. \ref{fig:excitation_ionization}). 
GHZ2 also lies in the region occupied by the five star-forming galaxies at $z>5$ (with no evidence of AGN at R$\sim1000$) studied by \citet{trump23} in the SMACS 0723 Early Release Observations \citep{pontoppidan22}, confirming that typical star-forming galaxies in the EoR have nebular properties and ISM conditions resembling those of AGNs at $5<z<7$, as also indicated by our modeling. 

Overall, these results indicate that even combining NIRSpec and MIRI, that is, considering the full rest-frame UV and optical wavelength coverage, we are not able to pin down the nature of the ionizing source in GHZ2.  
However, regardless of its physical nature, our modeling suggests in both cases the presence in GHZ2 of very low metallicity gas (Z$_{gas}$ $\sim 0.05$-$0.1$ Z$_\odot$), with high electron density and high ionization parameter.

\subsection{Ionization properties and SFR surface density}\label{sec:sigmaSFR_O32}

\begin{figure*}[t!]
\centering
\includegraphics[angle=0,width=0.97\linewidth,trim={0cm 0cm 0.cm 0cm},clip]{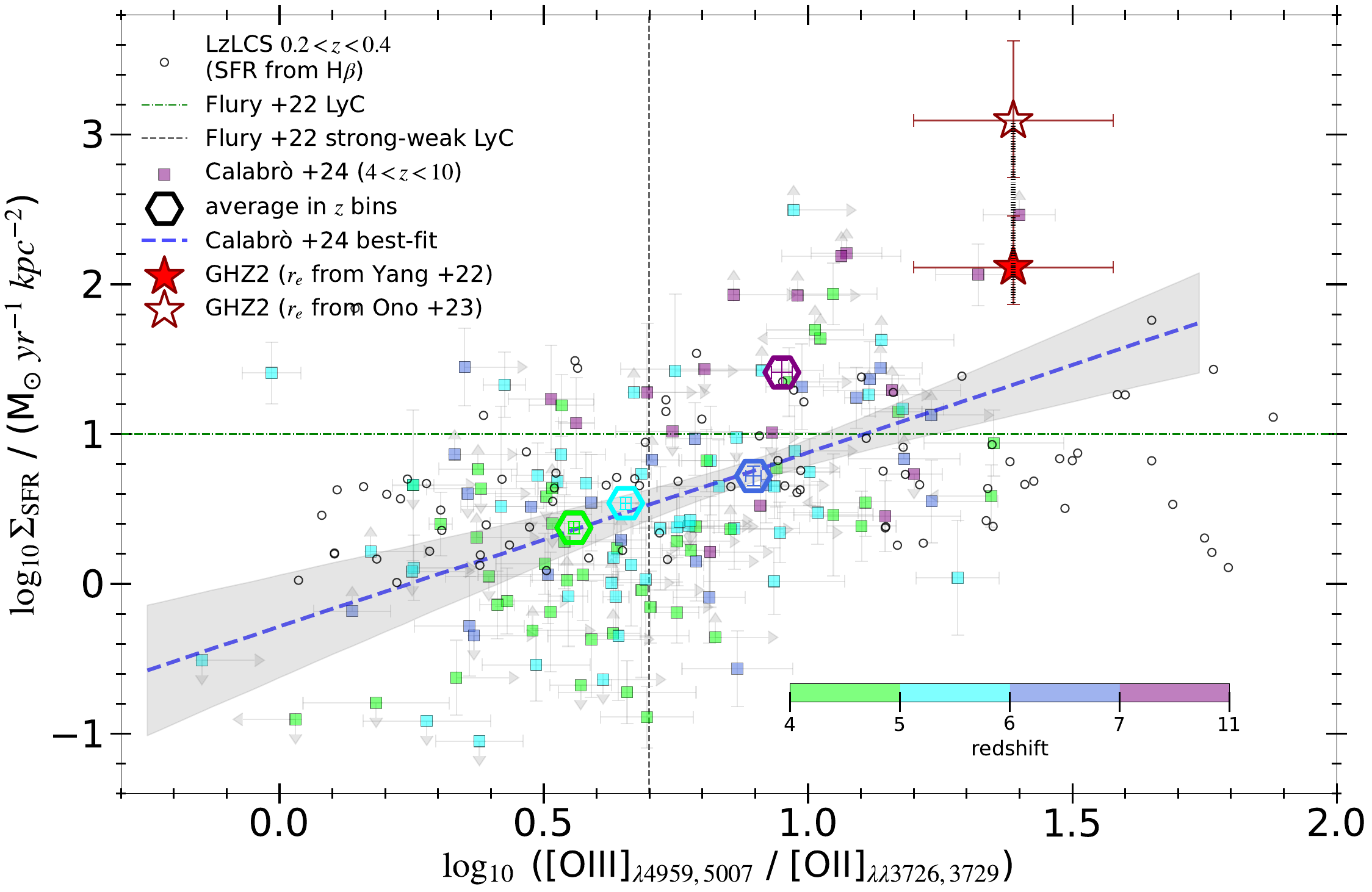}
\vspace{-0.01cm}
\caption{The \sigmaSFR vs O32 diagram. The position of GHZ2 is highlighted with a big red star. The sample of NIRSpec galaxies at redshifts $4<z<10$ from \citet{calabro24} is added with color coding based on redshift. The vertical and horizontal dashed lines represent the thresholds of \sigmaSFR and O32 for being a strong LyC leaker candidate, as obtained from the analysis of $66$ LyC leakers at $0.2<z<0.4$ (black empty circles) by \citet{flury22b_tables}. According these criteria, GHZ2 would be a strong LyC leaking candidate. 
} 
\label{fig:O32_sigmaSFR}
\end{figure*} 

The evolution of $\log_{10}$(U) across cosmic time can be explained as an effect of the increase of gas density and SFR surface density \citep[e.g.,][]{reddy23}, meaning that denser and more compact molecular clouds, hence elevated \sigmaSFR in high redshift galaxies, might be directly responsible for the increase of the ionization parameter, playing a more important role than metallicity. Significant correlations were indeed found between $n_e$, \sigmaSFR, and O32 \citep{shimakawa15,jiang19,reddy23,calabro24}, confirming the close relation between these quantities. 
According to hydrodynamical simulations \citep[e.g.,][]{ma16,sharma17}, enhanced \sigmaSFR and O32 also favor strong stellar feedback and outflows, which can carve channels in the ISM of the galaxies for the leakage of Lyman continuum radiation. Previous observations have reported a relation between \sigmaSFR and the escape fraction \fesc of ionizing photons \citep{heckman01,naidu20,flury22a_LyCdiagnostics_correlations}, showing that galaxies with higher \sigmaSFR have larger fractions of Lyman Continuum (LyC) leakers and higher \fesc. Similarly, \citet{izotov20} empirically found a close relation between O32 and the measured \fesc at low redshift. \citet{flury22a_LyCdiagnostics_correlations} put the two quantities together, establishing that galaxies with \sigmaSFR $> 10$ \msun yr$^{-1}$ kpc$^{-2}$ and O32 $>0.7$ are strong LyC leaker candidates, based on direct measurements of \fesc in a very large sample of local LyC leaking galaxies from the Low Redshift Lyman Continuum survey (LZLcS). 
We can investigate the properties of GHZ2 in the O32 vs \sigmaSFR diagram, and infer some information on the escape fraction. 

In Fig. \ref{fig:O32_sigmaSFR}, we compare GHZ2 to NIRSpec galaxies at $4<z<10$ and to the LyC leaking galaxies at $0.2<z<0.4$ observed by \citet{flury22b_tables}. 
We can see that GHZ2 lies in the upper right corner of the diagram, with \sigmaSFR and O32 significantly higher compared to typical values found in lower redshift galaxies. 
In detail, GHZ2 lies at the highest values of O32 and \sigmaSFR found for NIRSpec galaxies from \citet{calabro24}, and above the \sigmaSFR - O32 best-fit relation derived in that work. 
Considering the smaller size reported in the literature by \citet{ono23} would make this galaxy even more extreme, surpassing in \sigmaSFR and O32 all the sources observed at the EoR by the previous study.  

The position of GHZ2 in Fig. \ref{fig:O32_sigmaSFR} implies the galaxy is quite likely to leak LyC emission, satisfying all the conditions defined by \citet{flury22a_LyCdiagnostics_correlations}.  
Using the \fesc relation by \citet{mascia23}, calibrated on LyC leaking galaxies from the LZLcS, we derive an indirect estimate of \fesc $=0.10_{-0.01}^{+0.02}$. Alternatively, using the $\beta$ slope based relation by \citet{chisholm22} yields a very consistent value of $0.11_{-0.02}^{+0.02}$. 
This average \fesc of $0.10$ inferred for GHZ2 is consistent with that predicted for its \sigmaSFR value, using the \sigmaSFR - \fesc best-fit relation at $4<z<10$ by \citet{calabro24}.  
Moreover, in the NIRSpec spectrum there is a $2 \sigma$ flux excess at the position of MgII$_{\lambda2800}$, which needs confirmation. If confirmed through deeper observations, this would also be a typical feature of LyC leakers \citep{chisholm20}. 

All this suggests that GHZ2 could have a strong Lyman continuum output.  
Nevertheless, these findings also suggest that comparing with low-redshift analogs may be a limitation, possibly due to the extremely high mass density and SFR density of GHZ2. Indeed, while some of the galaxies from \citet{flury22b_tables} have O32 similar to GHZ2, none of them reaches the compact star-formation activity seen in this galaxy at $z=12.34$. 
We also note that the CIV$_{\lambda 1549}$/CIII]$_{\lambda\lambda 1907,1909}$ ratio (another Lymam continuum leaker indicator) was measured for GHZ2 by C24, and its value of $\simeq 3$ exceeds those typically found in strong LyC leakers at $z\sim0.3$-$0.4$ \citep{schaerer22a_CIV}.

%
Using this indirect derivation of \fesc, we can recalculate the ionizing photon production efficiency $\xi_{\rm ion}$ compared to the estimation in Z24, which assumes \fesc $=0$. Assuming negligible dust attenuation and the M$_{UV}$ from the companion papers, we can use the following formulation of \citet{schaerer16} as
\begin{equation}\label{xi_ion1}
  N_{\rm Lyc} [s^{-1}] = 2.1 \times 10^{12}  (1- \fesc)^{-1} L(\Hb) [erg s^{-1}]
\end{equation}
and 
\begin{equation}\label{xi_ion2}
\log(\xi_{\rm ion}) = \log(N_{\rm Lyc}) + 0.4 \times M_{\rm  UV} - 20.64,
\end{equation} 
where N$_{\rm Lyc}$ is the number of ionizing photons produced per unit time, obtaining $\log(\xi_{\rm ion}) = 25.72_{-0.15}^{+0.35}$, which is consistent with the lower limit of $25.3$ derived in Z24. This is comparable to values typically measured in Ly$\alpha$ emitters \citep{harikane18,sobral19}, and in $z > 7$ galaxies with strong [OIII] and UV line emission \citep{stark17,endsley21}. In particular, our result is very similar to the $\log(\xi_{\rm ion}) = 25.69$ measured in a strong CIV emitter at $z =7.045$ by \citet{stark15}. 
A significant correlation was also found between $\xi_{\rm ion}$, \mstar, and \sigmaSFR \citep{castellano23b}, with $\log(\xi_{\rm ion}) > 25$ found in galaxies with \sigmaSFR $>10$ \msun/yr/kpc$^2$. The $\xi_{\rm ion}$ estimated for GHZ2 is thus consistent with that expected for a system with extreme \sigmaSFR. 

The absence of any Ly$\alpha$ emission in GHZ2 
might give additional insights on the environment of this galaxy. 
Despite the relatively bright magnitude and escape fraction, the age of the source is too short and therefore there was not enough time to create a large enough ionized bubble for the Ly$\alpha$ to become visible.
Deeper observations at higher resolution in the future could put tighter constraints on the presence of faint, highly redshifted Ly$\alpha$. 


\subsection{Gas-phase metallicity from strong-line ratios}\label{sec:metallicity}


An estimate of the metallicity of GHZ2 was already made in C24 and Z24, using the emission line indicators available in their spectral range. In the former, they derived Z from the Ne3O2 index, the CIII]$_{\lambda 1909}$/OIII]$_{\lambda 1663}$ ratio, and the EW of CIII]. In the latter instead they estimated the ISM metal content from the R3 index. Adopting a variety of calibrations available in the literature, they suggest that the metallicity of GHZ2 is in the range between $\sim 3\%$ and $\sim 12\%$ solar. 

We now complement previous measurements of Z$_{gas}$ with additional diagnostics that exploit the full NIRSpec + MIRI coverage, including in particular the R2, R23, and O32 indices. 
R23 is one of the most widely adopted since, including both of the main ionization stages of oxygen (O$^+$ and O$^{++}$), it is not significantly affected by the ionization structure of the HII regions.  
However, it has a bi-valued metallicity solution, peaking at $12+\log$(O/H) $\simeq 8.0$ and decreasing toward higher and lower metallicities \citep{maiolino08}, which makes the dependence weak at the turnaround. R2 also has a quadratic behaviour as a function of metallicity, but it peaks at higher metallicity ($12+\log$(O/H) $\simeq 8.7$), hence it has more constraining power at low Z. Finally, the dependence of the O32 index on metallicity is mostly secondary as O32 depends on both Z and the ionization parameter, but it has the advantage of showing in general a monothonic increase toward lower metallicities. 
The combination of these three metallicity indicators with the other optical-based ones analyzed in C24 and Z24, can more tightly constrain the oxygen abundance of the galaxy, minimizing degeneracies and secondary dependences on other elements, as suggested by \citet{curti20}.  

To perform the calculations, we consider multiple relations, calibrated with the direct (T$_e$-based) method on galaxies from $z=0$ to higher redshifts for which the \OIII\ auroral line at $4363$\AA\ is available.  
In particular, we consider the strong-line calibrations derived by \citet{curti17,curti20} from SDSS galaxies, those derived by \citet{bian18} using a BPT-selected (high ionization and high excitation) sample from the SDSS, and those proposed by \citet{nakajima22} using a sample of local extremely metal poor galaxies (EMPGs). In addition, we also adopt the calibrations obtained by \citet{sanders24} by adding $16$ high redshift galaxies with auroral line detection recently observed by JWST to previous \OIII$_{\lambda 4363}$ detected samples observed from the ground. 
The results obtained with all these estimators are summarized in Table \ref{tab:metallicity}, and they range from a minimum of $4\%$ solar to a maximum of $12\%$ solar.  
In all cases, the $1 \sigma$ lower and upper errors on Z are derived by considering only the uncertainty of the line ratios. We also note that at $z>10$ we expect the metal content to be significantly lower than $0.3 \times$ Z$_\odot$ \citep{ucci23}, hence we can safely assume that we are in the low metallicity branch of the R2 and R23 relations. 

\begin{deluxetable}{cccccc}\label{linestable}
\tablecaption{Line indices measured for GHZ2}
\tablewidth{0pt}
\tabletypesize{\footnotesize}
\tablehead{
\colhead{\hspace{-.5cm}Index}\hspace{-.5cm} & \colhead{\hspace{-.5cm}Line ratio}\hspace{-.5cm} & \colhead{\hspace{-.5cm}Z$^\text{A}$}\hspace{-.5cm} & \colhead{\hspace{-.5cm}Z$^\text{B}$}\hspace{-.5cm} & \colhead{Z$^\text{C}$} & \colhead{Z$^\text{D}$} \\
\colhead{} & \colhead{} & \colhead{Z$_\odot$} & \colhead{Z$_\odot$} & \colhead{Z$_\odot$} & \colhead{Z$_\odot$}
}
\startdata       
R2 &  $-0.52 \pm 0.20$       &        -               &         -       &    $0.06_{0.03}^{0.11}$  &  $0.06_{0.04}^{0.09}$ \\
R3$^1$ & $0.72 \pm 0.11$     &        -               &   $0.12_{0.09}$ &    $0.08_{0.05}^{0.16}$  &  $0.05_{0.03}^{0.12}$ \\
R23 & $0.88 \pm 0.12$        &        -               &   $0.11_{0.06}$ &    $0.08_{0.05}^{0.19}$  &  $0.05_{0.03}^{0.13}$ \\
O32 & $1.39 \pm 0.19$        &  $0.11_{0.07}^{0.14}$  &       -         &            -             &  $0.07_{0.05}^{0.10}$ \\
Ne3O2$^2$ & $0.37 \pm 0.18$  &  $0.07_{0.06}^{0.10}$  &       -         &            -             &  $0.04_{0.02}^{0.05}$ \\
\footnotesize{Ne3O2Hd} & $0.59 \pm 0.12$    &  -      &      -         &            -             &  -     \\
&     &     &     &      &  \\
\hline
average &                    &  $0.08 \pm 0.02$       &      $0.11 \pm 0.04$    &      $0.07 \pm 0.03$     &   $0.054 $   \\
     &     &     &     &      &  $ \pm 0.009$  \\
\enddata
\tablecomments{Table with gas-phase metallicities (in units of Z$_\odot$) estimated for GHZ2 using multiple indices (first row) and the following calibrations from: \textbf{A :} \citet{bian18}, \textbf{B :} \citet{curti20}, \textbf{C :} \citet{nakajima22}, \textbf{D :} \citet{sanders24}. The average value from all the indices available for the same calibration set is shown in the last row. A metallicity of $0.06^{+0.04}_{-0.02}$ is obtained with the T$_e$ method from the OIII]$_{\lambda 1666}$/[OIII]$_{\lambda 5007}$\ line ratio. The R3, R23, O32, Ne3O2, and Ne3O2Hd indices are used to study the excitation and ionization properties in Section \ref{sec:R23_O32} and \ref{sec:OHNO}. $^1=$ see also \citet{zavala24}, $^2=$ see also \citet{castellano24}. }
\end{deluxetable}\label{tab:metallicity}

We find that, within the same calibration set, the various indices tend to give metallicities that are consistent among each other, suggesting that using only one index correctly informs the metallicity of the galaxy. A weighted average from all the available line ratios is also provided in Table \ref{tab:metallicity} for each calibration set. 
In some cases the metallicities could not be derived, so we did not consider them in the weighted average. For example, in \citet{bian18}, the R3 and R23 indices do not cover the lower branch, while adopting \citet{curti20} and \citet{nakajima22}, our O32 and Ne3O2 measurements fall outside of the range covered by their calibration. We note that in the calibrations by \citet{nakajima22}, the O32 and Ne3O2 indices show a quadratic, bi-valued behaviour at low metallicity, similar to that seen for the R23 index at high Z, with a narrow allowed range for the two line ratios, hence they would not be very informative on the chemical content of the galaxy. For these calibrations we consider the relations valid for galaxies with high EW(\Hb) $>200$ \AA. Even though this line and the underlying continuum are not detected with MIRI, the high O32 line ratio strongly suggests that we are in the high EW(\Hb) case, with expected values much higher that $200$ \AA\ according to the correlation shown in Fig. 22 of \citet{flury22a_LyCdiagnostics_correlations}. 

Most of the discrepancy within the metallicity range estimated for GHZ2 comes from applying different calibration sets, as also noted in the two companion papers from individual indices. This is likely due to the different sample selection, including different redshifts and (possibly) different intrinsic physical properties of the galaxies considered for the calibrations. On average, the relations from \citet{sanders24} tend to give metallicities that are slightly lower by $\sim1$-$2\%$ compared to the median value from all the methods, while \citet{curti20} tend to give slightly higher values. 
This variance (although small) of metallicity estimates highlights the challenge of finding local samples that truly reflect the characteristics of high-redshift galaxies, particularly those more extreme at $z>10$, as also discussed in previous sections.

Taking the weigthed averages together, the different calibrations suggest that the metallicity of GHZ2 is comprised between $5\%$ and $11\%$ solar, consistent with the previous estimation in C24 and Z24. If we further average these results we obtain a metallicity of $0.06_{-0.01}^{+0.02}$ Z$_\odot$ 
This is slightly above but also consistent with the metallicity of $0.04_{0.02}^{0.07}$ derived by C24 from SED fitting. 

The metallicity of GHZ2 is very low, although not significantly different from some extremely metal poor galaxies discovered at the EoR \citep{schaerer22b_JWST,arellano22,brinchmann23,rhoads23,nakajima23,morishita23,langeroodi23}, or in very low-mass galaxies at low and intermediate redshift \citep[e.g.,][]{calabro17,amorin12,amorin15,amorin17}. 
This result also suggests that GHZ2, despite the very young age of the Universe at $z\sim12$, already contains chemically enriched gas. 

\subsection{Electron temperature and T$_e$-based metallicity}\label{sec:electron_temperature}

Compared to our companion papers (C24 and Z24), the combination of NIRSpec and MIRI allows us to estimate the electron temperature of the gas from the OIII]$_{\lambda 1666}$/[OIII]$_{\lambda 5007}$ observed line ratio of $= 0.09 \pm 0.02$, and, consequently, the gas-phase metallicity using the direct (\te-based) method.

For the calculation of the metallicity, we use the python package \texttt{pyneb}. We first assume that the gas has an electron density of $10^3$ cm$^{-3}$, as suggested by the ALMA data and by the comparison with photoionization models. Then, we derive the electron temperature of O$^{2+}$, \te([O\iii]), from OIII]$_{\lambda 1666}$/[OIII]$_{\lambda 5007}$, adopting the collision strenghts from \citet{aggarwal99}. This yields \te([OIII]) $21200^{+2700}_{-2400}$~K.  
Regarding the temperature of O$^+$, \te([OII]), given the non detection of low-ionization auroral lines (e.g., [OII]$_{\lambda\lambda7322,7332}$), we infer it in an indirect way from \te([OIII]) using the relation of \citet{campbell86} :
\begin{equation}\label{eq:t2t3}
T_{\mathrm{e}}(\mathrm{O}~\textsc{ii}) = 0.7\times T_{\mathrm{e}}(\mathrm{O}~\textsc{iii}) + 3,000~\mathrm{K},
\end{equation}
as done by \citet{sanders24} for galaxies up to redshift $\sim 9$. 
However, we note that the results would not significantly change if we instead assume \te([OII]) $=$ \te([OIII]).

For the calculation of the oxygen abundance, we assume that O$^{3+}$ is negligible and that all O is in the form of O$^{2+}$ or O$^+$, which is valid also in galaxies with extremely high ionization conditions, as suggested by some recent works \citep[e.g.,][]{berg18,berg21}.
Using the electron temperatures estimated above, we finally derive the O$^{2+}$/H$^+$ and the O$^{+}$/H$\beta$ ratios from [OIII]$_{\lambda 5007}$/H$\beta$ and [OII]$_{\lambda 3727}$/H$\beta$, respectively. This yields an oxygen abundance (with the direct method) of $12+\log(O/H)=7.44^{+0.26}_{-0.24}$, that is, $0.06^{+0.04}_{-0.02}$ Z$_\odot$. This result is remarkably consistent with the metallicty derived by combining all the five strong line indicators (using the \Citealt{sanders24} calibration), as shown in Table \ref{tab:metallicity}, corroborating the very metal poor nature of GHZ2. 

\section{Conclusions}\label{sec:conclusions}

The combination of two instruments, NIRSpec and MIRI, in addition to the unique sensitivity of JWST in the near-infrared, has allowed us for the first time to probe the full rest-frame UV and optical spectral properties for a galaxy at $z>10$. 
The galaxy GHZ2/GLASSz-12, one of the brightest galaxies discovered during the first JWST imaging campaigns, and confirmed to have a spectroscopic redshift of $12.34$ (from C24 and Z24). 
Leveraging this unique dataset, we have investigated the ionization and metallicity properties of this galaxy in a comprehensive way, shedding light on the ISM conditions in pre-EoR. 

Combining six rest-frame optical indices, we have found that GHZ2 has very strong ionizing conditions as probed by extreme O32 and Ne3O2 line ratios (respectively $\sim 1.4$ and $\sim0.4$). These values are consistent with photoionization by either an AGN or star-formation, which we are not able to discriminate with the NIRSpec + MIRI spectra. 
Assuming photoionization from young massive stars, the line ratios suggest that the ionization parameter is very high between $\log$(U)$ = -2$ and $-1.5$, surpassing the average values found in star-forming galaxies at the EoR and in local analogs of high-redshift galaxies.
One possible underlying physical reason for these rather unique ionizing properties of GHZ2 is that the star-formation of the whole galaxy is still confined in a very compact radius. 
The very high \sigmaSFR ($\sim 10^2$ to $\sim 10^3$ \msun/yr/kpc$^2$) and \sigmaMass ($\sim 10^4$ to $\sim10^5$ \msun/pc$^2$), the high ionizing photon production efficiency ($\log(\xi_{\rm ion}) \simeq 25.7$), and the high gas densities ($10^3$ cm$^{-3}$ $<n_e<$ $10^4$ cm$^{-3}$) suggested by the nebular line ratios (also including ALMA data) are additional indicators of an extreme star-formation scenario that is not typical of galaxies at lower redshifts. A similarly high density of SF and \mstar\ has been found only in very few systems at $z<10$, such as in an extremely dense UV-bright starburst at $z=3.6$ analysed by Marques-Chaves et al. (2022), and in \rxc, which shows even more extreme \sigmaSFR ($>10^4$ \msun/yr/kpc$^2$) and ionizing conditions than GHZ2 \citep{topping24}.

These properties are in agreement with a scenario in which the galaxy has recently undergone a phase of elevated, very compact star-formation activity that has very rapidly enriched the ISM metallicity at the level of $\sim 5$-$10\%$ solar. 
This physical configuration also suggests that GHZ2 might be among the first promising candidates to start reionizing and polluting the surrounding intergalactic medium (IGM) before the EoR, possibly through radiatively driven winds from young massive stars in the recent past, as suggested by some recent models \citep{ferrara23a,ferrara23b}. 
The negligible dust attenuation (A$_V$ $\leq 0.1$ mag) estimated for GHZ2 in C24 and Z24 is consistent with a dust clearing outflow scenario outlined by those models. 
In agreement with this scenario, also the escape fraction of $0.1$, estimated from a variety of indirect indicators, indicate that the galaxy may have created channels through which ionizing photons can leak. 
The lack of Ly$\alpha$ suggests that GHZ2 is rather isolated and has not yet created a large ionized bubble required for that line to transmit in a completely neutral IGM expected at $z\simeq 12$. 
We also note that, even though the specific SFR (sSFR) of $\log$(sSFR/yr$^{-1}$)$=-8.10_{-0.27}^{+0.54}$ is currently below the super-Eddington threshold proposed by \citet{ferrara23b}, it might exceed that limit according to alternative \mstar estimations by Z24, or it might have reached those conditions in the past, depending on the exact shape of the star-formation history.  

The combination of the NIRSpec and MIRI spectra for GHZ2 provides an unique insight to investigate the physical properties of galaxies at $z>10$, which has become the next challenge for JWST.  
First, the UV to optical combined diagnostics yield ionization and metallicity properties that are in agreement with those estimated with UV-only or optical-only indicators. This suggests that analysis of a limited region of the spectrum, such as that still accessible with NIRSpec at very high-redshift, provide reliable results and, at the same time, an effective way to characterize statistically meaningful samples of objects in the pre-reionization cosmic phase. 
Nevertheless, MIRI could be used on the brightest galaxies to further test the consistency of the UV and optical diagnostics. 

Secondly, despite the large wavelength coverage of the NIRSpec + MIRI combined observations, UV+optical indicators still fail to unambigously assess the nature of the ionizing source, as they cannot discriminate between an AGN and star-formation in a low metallicity, high density, and extreme ionization environment. Deeper observations at high resolution targeting broad components of the CIV and Balmer lines, and fainter high-ionization lines in the UV to optical range are likely required if we want to clarify the ionizing mechanism of the earliest galaxies.

Finally, the extreme emission line properties and ionizing conditions of GHZ2 when compared to galaxies at lower redshifts suggests that the consolidated approach of using nebular line ratios of local, low mass, metal poor systems to infer the properties of high-z galaxies might have reached its limits when coming to extreme redshifts ($z \gtrsim 10$). This is due to the extremely dense ISM and star-formation that cannot be fully tested in local counterparts. Furthermore, galaxies at $z > 10$ are likely surrounded by an IGM that is still completely neutral and pristine, highlighting the stark differences in properties and environment between the two populations.
The challenge of finding samples of local analogs that are truly representative of the properties of high-redshift galaxies is evident in the slight variance observed among various metallicity calibrations. 
Direct observations of $z>10$ galaxies are thus necessary.

\begin{acknowledgments}
This work is based on observations made with the NASA/ESA/CSA {\it James Webb Space Telescope (JWST)}. The data presented in this article were obtained from the Mikulski Archive for Space Telescopes (MAST) at the Space Telescope Science Institute, which is operated by the Association of Universities for Research in Astronomy, Inc., under NASA contract NAS 5-03127 for {\it JWST}. The specific observations analyzed can be accessed via \dataset[doi: 10.17909/4r6b-bx96]{https://doi.org/10.17909/4r6b-bx96} and \dataset[doi: 10.17909/gret-mk52]{https://doi.org/10.17909/gret-mk52}. These observations are associated, respectively, with program JWST-GO-3073 and JWST-GO-3703.  
We acknowledge financial support from NASA through grant JWST-ERS-1342. Support was also provided by INAF Mini-grant ``Reionization and Fundamental Cosmology with High-Redshift Galaxies. AC acknowledges support from the INAF Large Grant for Extragalactic Surveys with JWST. We acknowledge support from the PRIN 2022 MUR project 2022CB3PJ3 - First Light And Galaxy aSsembly (FLAGS) funded by the European Union – Next Generation EU." 

\end{acknowledgments}

%

\vspace{5mm}

{}



\end{document}